\documentclass[prb,twocolumn,showpacs,preprintnumbers,amsmath,amssymb]{revtex4}

\usepackage{graphicx}
\usepackage{dcolumn}
\usepackage{bm}

\begin{document}

\title{ Polarisation-sensitive terahertz detection by multicontact photoconductive receivers }
\author{E. Castro-Camus}
\email{e.castro-camus1@physics.ox.ac.uk}
\author{J. Lloyd-Hughes}
\affiliation{University of Oxford, Department of Physics,
Clarendon Laboratory, Parks Road, Oxford OX1 3PU, United Kingdom}
\author{M.D. Fraser}
\author{H.H. Tan}
\author{C. Jagadish}
\affiliation{ Department of Electronic Materials Engineering,
Research School of Physical Sciences and Engineering, Institute of
Advanced Studies, Australian National University, Canberra ACT 0200,
Australia }
\author{M.B. Johnston}
\affiliation{University of Oxford, Department of Physics,
Clarendon Laboratory, Parks Road, Oxford OX1 3PU, United Kingdom}

\parbox{17.2cm}{\centering{\footnotesize APPLIED PHYSICS LETTERS {\bf 86}, 254102 (2005)}}

 \received{10 March 2005} \accepted{18 May 2005}

\begin{abstract}
We have developed a terahertz radiation detector that measures both
the amplitude and polarisation of the electric field as a function
of time. The device is a three-contact photoconductive receiver
designed so that two orthogonal electric field components of an
arbitrary polarised electromagnetic wave may be detected
simultaneously. The detector was fabricated on Fe$^{+}$
ion-implanted InP. Polarisation-sensitive detection is demonstrated
with an extinction ratio better than 100:1. This type of device will
have immediate application in studies of birefringent and optically
active materials in the far-infrared region of the spectrum.
\end{abstract}

\pacs{07.50.-e, 07.57.-c, 07.57.Kp , 07.57.Pt, 07.60.Fs, 42.25.Ja, 71.55.Eq , 78.20.Ek, 78.20.Fm}
\keywords{Terahertz, photoconductive, receiver, antenna, optical
activity, circular dichroism}
\maketitle

The far-infrared, or terahertz (THz), region of the
electromagnetic spectrum encompasses the energy range of many
collective processes in condensed matter physics and
macromolecular chemistry. However, in the past this spectral
region has been relatively unexplored owing to a lack of bright
radiation sources and appropriate detectors. The technique of THz
time domain spectroscopy (TDS),\cite{AustonC85,SmithAN88} which
has developed rapidly as a result of advances in ultra-short
pulsed laser technology, now provides a very sensitive probe
across the THz band. TDS is currently an invaluable tool in
condensed matter physics \cite{HuberTBBAL01, LeitenstorferHTBBA02,
KaindlHCLC03} and macromolecular
chemistry.\cite{cpl03,schmuttenmaer2004}

To date THz-TDS techniques have relied on linearly polarised
emitters and detectors.  However, for spectroscopy of birefringent
and optically active materials it is also important to measure the
polarisation state of radiation before and after it has interacted
with the material. Here we report on a detector that enables such a
THz-TDS system to be realised.

Vibrational circular dichroism (VCD) spectroscopy is a new technique
which has substantial potential in the fields of macromolecular
chemistry and structural biology.\cite{nafie96}  Akin to the
established technique of (ultraviolet) circular dichroism, VCD is
used to analyse the structure of chiral molecules. It is predicted
that VCD will be more powerful than conventional circular dichroism
for stereo-chemical structure determination.\cite{nafie96} However
the technique is currently limited by insensitive and narrow band
spectrometers.

Of particular interest to biochemists is the structure and function
of proteins and nucleic acids. These chiral biomolecules have
vibrational and librational modes in the THz region and the THz
optical activity of these modes are starting to be studied
experimentally.\cite{xu19,XuRGSSBAP03}  THz frequency VCD is already
finding application in fields as distinct as biochemical research
\cite{salzman:2175} and astrobiology.\cite{XuRGSSBAP03} In the
future the ability to perform VCD using a polarisation sensitive
THz-TDS technique should enhance the bandwidth and sensitivity of
measurements, and allow dynamic time-resolved studies to be
performed.

In order to perform polarisation sensitive THz-TDS, it is necessary
to be able to measure two (preferably orthogonal) electric field
components of a terahertz transient. Theoretically it is possible to
do this using a conventional (two contact) photoconductive receiver.
That is, measure one electric field component and then rotate the
receiver by $90^{\circ}$ and measure the other component. However in
practice this procedure has two major disadvantages: Firstly, during
the rotation of the photoconductive receiver, any slight
misalignment will significantly shift the relative phase of the
electric field components; secondly the data acquisition time is
increased as both components are recorded separately.  In order to
avoid these disadvantages an integrated receiver capable of
measuring both components simultaneously is needed. Such a detector
may be realised by fabricating a three-contact photoconductive
receiver.

\begin{figure}
\begin{center}
\includegraphics[width=8.0cm]{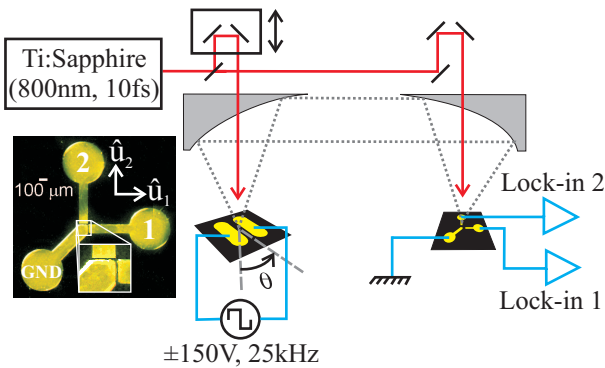}
\end{center}
\caption{ (Colour online) Diagram of experimental apparatus used for
simultaneous detection of horizontal and vertical components of the
electric field of a THz transient. A SI-GaAs photoconductive switch
was used as emitter, parabolic mirrors were used to collect and
focus the THz radiation onto the three-contact photoconductive
receiver. One of the contacts was used as common (GND) and the other
two were amplified independently to obtain the two orthogonal
components. Inset: Photograph of a three-contact photoconductive
receiver structure formed by two $16\mu \rm{m}$ gaps in orthogonal
directions in order to measure the perpendicular components of the
THz electric field. The unit vectors $\mathbf{\hat{u}}_1$ and
$\mathbf{\hat{u}}_2$ represent the directions between the earth
contact and contacts 1 and 2 respectively.  The photograph was taken
using an optical microscope. } \label{f:Photo}
\end{figure}

The three-contact receiver we developed is shown in
Fig.~\ref{f:Photo}. When designing the three-contact receiver we
considered two main constraints. Firstly, the unit vectors
($\mathbf{\hat{u}}_1$ and $\mathbf{\hat{u}}_2$) normal to the gaps
formed between the earth contact and the other two contacts (1 and
2) need to be orthogonal. Secondly it is necessary that both gaps
are within an area smaller than the laser beam waist and the focus
spot size of the THz radiation (a circle of radius
$\sim100\,\rm{\mu m}$).  This last condition is necessary in order
to have uniform laser and THz illumination across both gap
regions.

The performance of a photoconductive receiver depends strongly on
the material from which the device is fabricated.  Material
dependent carrier trapping and recombination times play an essential
role in photoconductive receiver device performance. Specifically,
long carrier lifetimes will permit the reception of large amounts of
noise and short carrier lifetimes will reduce the signal level and
accuracy. Modified semiconductor materials such as
low-temperature-grown or ion-implanted GaAs/InP are typically used,
as carrier trapping times in these materials may be
controlled.\cite{liu04,ShenUBLDGBTE04}

In order to fabricate the three-contact PCS device (shown in
Fig.~\ref{f:Photo}) we have implanted semi-insulating InP (100)
substrates using $2.0\,\rm{MeV}$ and $0.8\,\rm{MeV}$ $\rm{Fe^+}$
ions with doses of $1.0\times10^{13}\,\rm{cm^{-2}}$ and
$2.5\times10^{12}\,\rm{cm^{-2}}$ respectively.  These multi-energy
implants gave an approximately uniform density of vacancies to a
depth of 1\,$\mu$m, resulting in a carrier lifetime of
$\sim130\,\rm{fs}$.\cite{carmody:1074} The samples were subsequently
annealed at $500^{\circ}\rm{C}$ for 30 minutes under a PH$_3$
atmosphere. Finally the electrodes were defined using standard
photolithography and lift-off techniques. The Cr/Au contacts were
deposited to a thickness of 20/250\,nm using a thermal evaporator.

In order to measure a THz electric field $\mathbf{E}_{\rm THz}$
using a photoconductive receiver it is necessary to gate the
receiver with an ultra-short laser pulse.  The laser pulse
generates free charge carriers in the semiconductor substrate,
which are accelerated by $\mathbf{E}_{\rm THz}$ thus generating a
current $I$ between two contacts.  Assuming a laser pulse of the
form $\rm{sech}^2(1.76t/t_0)$ where $t_0$ is the
full-width-at-half-maximum, the current measured through contact
$i$ in the photoconductive receiver described here, is related to
$\mathbf{E}_{\rm THz}$ by\cite{kono:898}:
\begin{widetext}
\begin{equation}
{ I_{i}(t) \propto \int_{-\infty}^\infty \mathbf{E}_{\rm THz}(t')
\cdot \mathbf{\hat{u}}_i\, e^{-(t'-t)/\tau}  \times
[1+\tanh(1.76(t'-t)/ t_0)] \rm{d}t' } \label{e:Iconventional}
\end{equation}
\end{widetext}
 where
$\mathbf{E}_{\rm THz}(t')$ is the THz electric field,
$\mathbf{\hat{u}}_1$ and $\mathbf{\hat{u}}_2$ are unit vectors in
the direction of the two gaps between the contact ($i=$1,2) and
the earth electrode. $\tau$ is the lifetime of free carriers.

The three-contact photoconductive receiver was tested using the
setup shown in Fig.~\ref{f:Photo}. A linearly polarised THz
transient was generated by exciting a $400\,\rm{\mu m}$ gap SI-GaAs
photoconductive switch emitter biased by a $\pm150\,\rm{V}$ square
wave at a frequency of 25\,kHz. The emitted THz transients were
collected in the back reflection geometry and then focused on to the
receiver using off-axis parabolic mirrors. A Ti:Sapphire chirped
mirror oscillator with a 75\,MHz repetition rate provided 10\,fs
pulses of 4\,nJ and 800\,nm centre wavelength, which were used to
excite the emitter. A 0.4\,nJ fraction split from the original pulse
was used to gate the receiver.

Two separate lock-in amplifiers were used to record the currents
($I_1$ and $I_2$) through the two contacts. The lock-in amplifiers
and the common electrode of the receiver were connected to a
common earth, and the references of both lock-in amplifiers were
locked to a TTL signal provided by the 25\,kHz signal generator
(used to drive the THz emitter). In all measurements the
$I_{1}(t)$ signal from Lock-in 1 and $I_{2}(t)$ signal from
Lock-in 2 were recorded simultaneously at each time step using a
multichannel analogue to digital converter.

\begin{figure}
\begin{center}
\includegraphics[width=8.0cm]{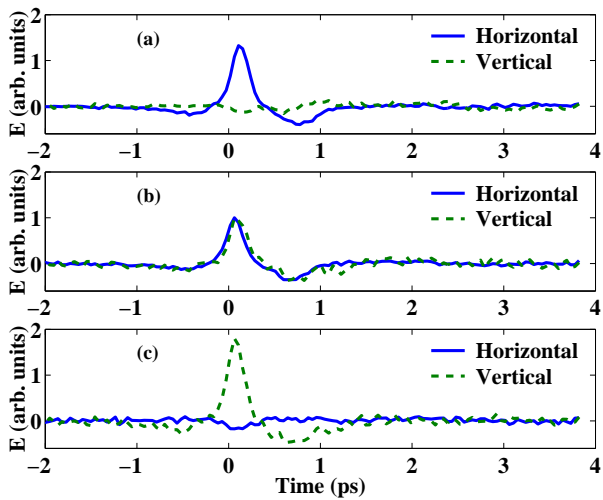}
\end{center}
\caption{(Colour online) Horizontal (solid) and vertical (dashed)
components of THz electric field (obtained form measured voltage)
are plotted against time with the emitter at (a) $0$, (b) $45$ and
(c) $90^\circ$ respectively.} \label{f:Lineal}
\end{figure}

The photoconductive switch emitter was mounted on a graduated
rotation stage that allowed the gap, and hence the polarisation of
the emitted THz transient, to be rotated. Both $I_1(t)$ and $I_2(t)$
where measured averaging over 90 scans at three different angles of
the emitter ($0$, $45$ and $90^{\circ}$).  The measurements were
taken in an evacuated chamber at a pressure of 25\,mbar to avoid
water vapor absorption.  The THz electric field was calculated by
differentiating numerically the two $I(t)$ traces measured by the
lock-in amplifiers according to Eq.~\ref{e:Iconventional}.  The
horizontal $E_{\rm{H}}$ and vertical $E_{\rm{V}}$ electric field
components are plotted against time in Figs.~\ref{f:Lineal} (a), (b)
and (c) for the emitter at angles $0^\circ$, $45^\circ$ and
$90^\circ$ respectively. The results demonstrate that the
three-contact photoconductive receiver acts as a polarisation
sensitive detector. The polarisation selectivity of the detector was
assessed by measuring the cross polarised extinction ratio. This
ratio was found to be 108:1 (128:1) for the horizontally
(vertically) oriented emitter. It should be noted that  the
polarisation of the radiation arriving at the detector may not be
perfectly linear, as photoconductive emitters do not produce purely
dipolar radiation.\cite{rudd01} Therefore, the true extinction ratio
of the detector may be higher.

\begin{figure}
\begin{center}
\includegraphics[width=8.0cm]{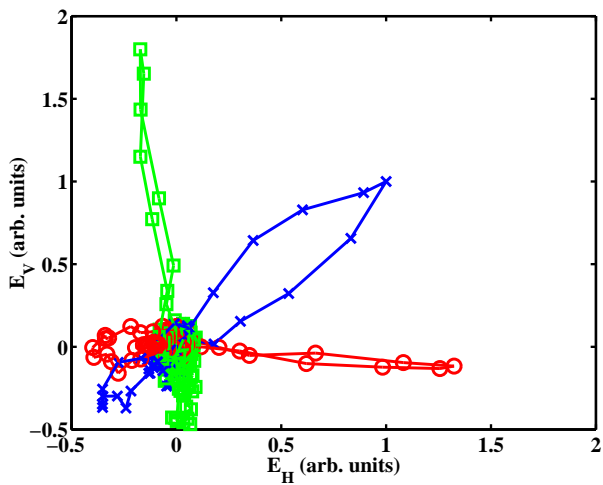}
\end{center}
\caption{ (Colour online) Parametric representation of horizontal
and vertical components of THz electric field for three emitter
orientations; $0$ (circles), $45$ (crosses) and $90^\circ$
(squares). The THz wavevector points out of the paper plane. In this
plot the angle of polarisation for the three waves is easily
observed. } \label{f:Parametric}
\end{figure}

In Fig.~\ref{f:Parametric} a parametric plot of the data shown in
Fig.~\ref{f:Lineal} is presented.  For an ideal linearly polarised
source the three sets of data should form straight lines at 0, 45
and 90$^{\circ}$ in the $E_{\rm{H}}-E_{\rm{V}}$ plane. However, the
measured angles of polarisation (from the horizontal plane) are
-5.5, 39, and 98$^{\circ}$ respectively and the polarisation appears
to be slightly elliptical (especially in the 45$^{\circ}$ case).
These discrepancies arise from a number of sources: It has been
shown previously that photoconductive switch emitters produce a
small quadrupole field leading to a cross polarised electric field
component.\cite{cai97,rudd01}  Furthermore, low \textit{f}-number
collection systems (such as the \textit{f}/1.5 system used in this
experiment) inevitably lead to linearly polarised radiation becoming
slightly elliptical.\cite{rudd01}

The signal-to-noise (SNR) ratio in our experiments was measured to
be 175:1 for the three contact receiver. We obtained a similar ratio
for a conventional two-contact ``bow-tie'' receiver, which we
fabricated on a piece of the same substrate material. This indicates
that the SNR performance of this device is limited by the substrate
material rather than the receiver design.  Therefore the device
sensitivity could be improved greatly by using optimised
ion-implanted InP or GaAs substrates. Indeed optimised low
temperature MBE-grown GaAs has been shown to have excellent SNR
performance\cite{ShenUBLDGBTE04} which should be replicated in a
three-contact device fabricated on that material.

In conclusion, the design of a novel integrated detector capable of
measuring both components of an arbitrarily polarised THz transient
was presented as well as experimental evidence of its effectiveness.
This integrated three-contact detector is expected to be very useful
for further studies of time-domain circular dichroism spectroscopy
and should have a wide range of applications in basic research and
industry.

The authors would like to thank the EPSRC (UK) and the Royal
Society for financial support of this work, ECC wishes to thank
CONACyT (M\'exico) for a scholarship. Australian authors would
like to acknowledge the financial support of the Australian
Research Council.

\bibliographystyle{apsrev}

\end{document}